\documentclass[12pt,preprint]{aastex}

\def\kms{\ifmmode{\rm km\thinspace s^{-1}}\else km\thinspace s$^{-1}$\fi} 

\def\H{HIP~50796}

\def\today{\number\year\space \ifcase\month\or  January\or February\or
        March\or April\or May\or June\or July\or August\or
September\or
        October\or November\or December\fi\space \number\day}

\slugcomment{To appear in The Astronomical Journal, January 2006 issue}

\shortauthors{Torres}
\shorttitle{\H}

\begin{document}

\title{The Astrometric-Spectroscopic Binary System \H:\break An
Overmassive Companion}


\author{Guillermo Torres}

\affil{Harvard-Smithsonian Center for Astrophysics, 60 Garden St.,
Cambridge, MA 02138}
\email{gtorres@cfa.harvard.edu}

\begin{abstract} 

We report spectroscopic observations of the star \H, previously
considered (but later rejected) as a candidate member of the TW~Hya
association. Our measurements reveal it to be a single-lined binary
with an orbital period of 570 days and an eccentricity of $e = 0.61$.
The astrometric signature of this orbit was previously detected by the
HIPPARCOS satellite in the form of curvature in the proper motion
components, although the period was unknown at the time. By combining
our radial velocity measurements with the HIPPARCOS intermediate data
(abscissae residuals) we are able to derive the full three-dimensional
orbit, and determine the dynamical mass of the unseen companion as
well as a revised trigonometric parallax that accounts for the orbital
motion.  Given our primary mass estimate of 0.73~M$_{\sun}$ (mid-K
dwarf), the companion mass is determined to be 0.89~M$_{\sun}$, or
$\sim$20\% \emph{larger} than the primary.  The likely explanation for
the larger mass without any apparent contribution to the light is that
the companion is itself a closer binary composed of M dwarfs. The
near-infrared excess and X-ray emission displayed by \H\ support this.
Our photometric modeling of the excess leads to a lower limit to the
mass ratio of the close binary of $q \sim 0.8$, and individual masses
of 0.44---\,0.48~M$_{\sun}$ and 0.41---\,0.44~M$_{\sun}$. The new
parallax ($\pi = 20.6 \pm 1.9$ mas) is significantly smaller than the
original HIPPARCOS value, and more precise. 
	
\end{abstract}

\keywords{binaries: spectroscopic --- binaries: visual ---
stars: individual (\H) --- stars: late-type --- techniques: radial
velocities --- techniques: spectroscopic}

\section{Introduction}
\label{sec:introduction}

In recent years there has been considerable interest in the study of
nearby groupings of young stars that do not appear to be associated
with any molecular clouds. The prototypical example is perhaps the
TW~Hydrae association \citep{Kastner:97}, with an estimated age of
8--10~Myr and some two dozen members. Searches for new members of this
group have relied on X-ray properties, kinematics, infrared (2MASS)
colors, and spectral features \citep[e.g.,][]{Makarov:01, Gizis:02,
Tachihara:03, Song:03}, and have produced many candidates and
motivated detailed follow-up studies to confirm them. 

One of those searches, by \cite{Makarov:01}, used proper motions from
the Tycho-2 catalog \citep{Hog:00} for X-ray sources listed in the
ROSAT Bright Source Catalog \citep{Voges:99}. The authors considered a
large area surrounding the previously known members of the TW~Hya
association, and proposed a list of 23 candidate members that were
subsequently followed up spectroscopically by \cite{Torres:01},
\cite{Song:02}, and \cite{Torres:03}. The present paper discusses one
of those objects, \H\ (also known as BD$-$09~3055, GSC~05493~00324,
1RXS~J102219.2$-$103302, TYC~5493--324--1, $\alpha = 10^{\rm h}
22^{\rm m} 17\fs99$, $\delta = -10\arcdeg 32\arcmin 15\farcs5$, J2000,
$V = 10.80$, SpT = K). The kinematic model of the association
constructed by \cite{Makarov:01} made specific predictions as to the
radial velocity expected for each candidate. Although the measurements
for \H\ by \cite{Torres:01} happened to agree perfectly with those
predictions ($+13.1 \pm 1.0$~\kms), \cite{Song:02} found that the star
shows no detectable Li~$\lambda$6708 absorption in its spectrum
(equivalent width $< 10$ m\AA), which indicates it is not very young
and therefore essentially rules it out as a true member.  They also
measured a radial velocity of $+22.4 \pm 0.9$~\kms\ that is nearly
10~\kms\ different from the previous measurements, and proposed that
the object is a spectroscopic binary. Further measurements by
\cite{Torres:03} confirmed the velocity variations and indicated an
orbital period of perhaps 1 or 2 years. 

Even though it was clear that \H\ is not related the TW~Hya
association, the object was kept on the observing list at the
Harvard-Smithsonian Center for Astrophysics (CfA) for the purpose of
establishing the orbit. Preliminary solutions appeared to imply a mass
for the companion that is larger than the primary, which was puzzling
given that the spectrum is single-lined. Additionally, the HIPPARCOS
satellite \citep{ESA:97} detected curvature in the proper motion
further supporting the binary nature of the object. We present here
the complete analysis of \H\ that combines the astrometry and the
spectroscopy and allows us to investigate in more detail the nature of
the overmassive secondary star. 

\section{Spectroscopic observations and reductions}
\label{sec:spectroscopy}

\H\ was observed spectroscopically at the CfA with an echelle
spectrograph on the 1.5-m Tillinghast reflector at the F.\ L.\ Whipple
observatory on Mt.\ Hopkins (Arizona).  A single echelle order was
recorded using a photon-counting intensified Reticon detector at a
central wavelength of 5187\,\AA, giving a spectral coverage of
45\,\AA.  The resolving power is $\lambda/\Delta\lambda\approx
35,\!000$. A total of 34 spectra were obtained from April 2002 until
May 2005, and the signal-to-noise (S/N) ratios achieved range from 5
to about 30 per resolution element of 8.5~\kms. Four archival spectra
(S/N of 9--30) were also used, and were obtained for another program
in 1985--1986 with nearly identical instrumentation on the Multiple
Mirror Telescope (also on Mt.\ Hopkins, Arizona), prior to its
conversion to a monolithic mirror.

Radial velocities were derived using {\tt XCSAO} \citep{Kurtz:98}, a
one-dimensional cross-correlation program well suited to our
relatively low S/N spectra that runs under IRAF\footnote{IRAF is
distributed by the National Optical Astronomy Observatories, which is
operated by the Association of Universities for Research in Astronomy,
Inc., under contract with the National Science Foundation.}. For the
template we used a synthetic spectrum selected from an extensive
library of spectra based on model atmospheres by R.\ L.\
Kurucz\footnote{Available at {\tt http://cfaku5.cfa.harvard.edu}.}
\citep[see][]{Nordstrom:94, Latham:02}. These calculated spectra are
available for a wide range of effective temperatures ($T_{\rm eff}$),
projected rotational velocities ($v \sin i$), surface gravities ($\log
g$) and metallicities. The optimum template was determined from grids
of cross-correlations over broad ranges in $T_{\rm eff}$ and $v \sin
i$, seeking to maximize the average correlation weighted by the
strength of each exposure. Solar metallicity was assumed to begin
with.  The above calculations were repeated for a range of $\log g$
values from 2.0 to 5.0, which allowed us to derive a rough estimate of
the surface gravity for \H.  We obtained $\log g = 4.2 \pm 0.5$ and
$T_{\rm eff} = 4600 \pm 150$~K.  The entire procedure was repeated for
metallicities different from solar, but the best match was achieved
for templates with the solar composition. The temperature we derive
corresponds to a main-sequence star with spectral type mid-K
\citep{Cox:00}. The rotational broadening was found to be negligible
(formally $v \sin i = 1 \pm 3$~\kms). \cite{Song:03} reported a
considerably larger value of 8~\kms. 

The stability of the zero-point of our velocity system was monitored
by means of exposures of the dusk and dawn sky, and small run-to-run
corrections were applied in the manner described by \cite{Latham:92}.
The radial velocities in the heliocentric frame including these
corrections are listed in Table~\ref{tab:rvs}. The typical precision
of a single measurement is 0.5~\kms.  Variations over the 19.4 years
of coverage are obvious and show a peak-to-peak amplitude of about
40~\kms. The period is 570 days, and the eccentricity ($e = 0.61$) is
quite significant. An orbital solution based on these velocities is
presented in the second column of Table~\ref{tab:elements}. Although
it is no longer relevant in connection with the possible membership to
the TW~Hya association, we mention in passing that the center-of-mass
velocity $\gamma = +24.77 \pm 0.14$~\kms\ of \H\ is far from the value
of $+13.1$~\kms\ that had been predicted by \cite{Makarov:01}. The
most significant finding, however, is the large mass function of the
binary: $f(M) = 0.258 \pm 0.011$.  For a typical primary mass
corresponding to the temperature we derive, the mass function implies
a secondary that is more massive than the primary, yet no signs of
another star are obvious in our spectra. We discuss this further
below. 

\section{Astrometric observations}
\label{sec:astrometry}

\H\ was observed by the HIPPARCOS satellite from December 1989 to
November 1992, during which 53 astrometric measurements were made. The
trigonometric parallax was determined to be $\pi_{\rm HIP} = 29.40 \pm
2.69$~mas. The uncertainty is somewhat larger than usual, but perhaps
understandably so given that the star is relatively faint ($V =
10.80$).  Additionally, the HIPPARCOS reductions revealed significant
curvature in the proper motion in the amount of $d\mu_{\alpha}/dt =
-10.90 \pm 6.51$ mas~yr$^{-2}$ and $d\mu_{\delta}/dt = +22.33 \pm
5.84$ mas~yr$^{-2}$, most likely a reflection of the orbital motion of
the object. Given its period of 570 days, however, nearly two cycles
of the orbit were covered during the HIPPARCOS campaign (which lasted
1058 days) so that the meaning of those coefficients is obscured.  The
proper motions and the parallax as reported in the HIPPARCOS catalog
are also likely to be compromised. 

Since the satellite observations are sensitive enough to have detected
the orbital motion on the plane of the sky, the best course of action
is therefore to make use of the individual astrometric measurements
\citep[``abscissae residuals";][]{ESA:97} together with the velocities
to constrain the orbit, and at the same time re-derive the position,
proper motion, and parallax. Those HIPPARCOS measurements are listed
in Table~\ref{tab:hip}.  We describe the global solution below. 

\section{Orbital solution}
\label{sec:orbit1}

The radial velocities allow for the determination of the usual
spectroscopic elements, which are the period ($P$), center-of-mass
velocity ($\gamma$), semi-amplitude of the velocity variation ($K$),
eccentricity ($e$), longitude of periastron for the primary
($\omega_1$), and time of periastron passage ($T$). On the other hand,
the visual elements constrained by the astrometry are the angular
semimajor axis ($a$), the inclination angle ($i$), the position angle
of the ascending node ($\Omega$), as well as $P$, $e$, $\omega_1$, and
$T$.  The combination of the two kinds of measurements thus provides
redundancy in the four latter orbital elements that strengthens the
solution.  Furthermore, the fact that we see no indication of the
secondary in the spectrum of \H\ means that the motion seen by
HIPPARCOS should essentially be that of the primary star around the
center of mass of the binary.  This introduces a relation between the
angular semimajor axis of the primary, $a_1$, and its velocity
amplitude, given by
 \begin{equation}
a_1 = 9.191967 \times 10^{-5}\cdot \pi K P \sqrt{1 - e^2}/ \sin i~,
\label{eq:a1}
 \end{equation}
 where $a_1$ and the trigonometric parallax $\pi$ are expressed in
milli-arc seconds, $K$ in $\kms$, and $P$ in days.  We use this
relation to eliminate the angular semimajor axis ($a_1$) as an
unknown.  Since the HIPPARCOS observations are made in an absolute
frame of reference, the abscissae residuals contain information on the
parallax as well as the position and proper motion of the barycenter
of the binary. Five additional variables are thus introduced
($\Delta\alpha^*$, $\Delta\delta$, $\Delta\mu_{\alpha}^*$,
$\Delta\mu_{\delta}$, $\Delta\pi$)\footnote{Following the practice of
the HIPPARCOS catalog we define $\Delta\alpha^* \equiv \Delta\alpha
\cos\delta$ and $\Delta\mu_{\alpha}^* \equiv \Delta\mu_{\alpha}
\cos\delta$.}, which represent corrections to a fiducial point of
reference that yield improved estimates of those quantities over the
catalog values \citep[see][]{ESA:97}. 

A total of 13 variables were adjusted in the solution, using standard
non-linear least-squares techniques \citep{Press:92}. The formalism
used for incorporating the abscissae residuals from HIPPARCOS into the
fit follows closely that described by \cite{vanLeeuwen:98} and
\cite{Pourbaix:00}, and is described in more detail in the Appendix.
The relative weights of the spectroscopic and astrometric observations
were adjusted to yield separate reduced $\chi^2$ values near unity. In
this way we determined a scale factor for the original uncertainties
of the HIPPARCOS abscissae residuals of 1.14, and a factor of 1.43 for
the original velocity errors. 

Due to the faintness of the target the median error of a single
HIPPARCOS measurement is 5.7 mas, so that the constraint provided by
the astrometry is relatively weak. Nevertheless, the results of this
initial fit yielded improvements in all of the orbital elements in
common with the spectroscopy. The value derived for the angular
semimajor axis of the primary was $a_1 = 16.9 \pm 1.6$~mas.  As a test
to see if the astrometry is able to detect the light contribution from
the companion, we ran another solution in which we left the semimajor
axis of the apparent orbit as a free parameter. In this case that
semimajor axis ($a_{\rm phot}$) corresponds to the motion of the
center of light of the binary, or photocenter, rather than that of the
primary, and it should in principle be smaller if the light from the
secondary is significant. This fit gave a value of $a_{\rm phot} =
18.1 \pm 3.6$~mas, which is not only not smaller but is also much more
uncertain (because it is no longer constrained by the spectroscopy
through eq.[\ref{eq:a1}]). Nevertheless, it is still consistent with
the previous result, within the errors.  We conclude from this that
the companion is not bright enough to produce a measurable astrometric
effect, and we proceed under the assumption that that star is
invisible (but see \S\ref{sec:orbit2}). 

Slight differences were found in the proper motion components
($\mu_{\alpha}^*$, $\mu_{\delta}$) compared to the values reported in
the HIPPARCOS catalog, which was not unexpected given that the catalog
values do not account for orbital motion. An external check on these
motions is available from the Tycho-2 catalog \citep{Hog:00}, in which
$\mu_{\alpha}^*$ and $\mu_{\delta}$ are derived by combining the
HIPPARCOS position (mean epoch 1991.25) with positions from
ground-based meridian-circle and photographic catalogs going back
several decades, and up to a century in some cases. Given that the
period of \H\ is only 570 days, the orbital motion over a baseline of
decades should tend to average out in the Tycho-2 analysis, resulting
in a more accurate measure of the proper motion than that reported by
HIPPARCOS. Indeed, as shown in Table~\ref{tab:pm} the values of
$\mu_{\alpha}^*$ and $\mu_{\delta}$ from our initial fit are much
closer to the Tycho-2 values than the HIPPARCOS values. This also
suggests that the solution might benefit if we made use of the
constraint provided by Tycho-2.  Consequently, we incorporated the
Tycho-2 proper motions into the fit as measurements, along with their
uncertainties.  The resulting proper motions from this combined
solution are listed for comparison in the fourth row of
Table~\ref{tab:pm}. The remaining orbital elements are hardly
affected, except for the slightly smaller uncertainties. 
	
The complete results of this combined fit are given in the third
column of Table~\ref{tab:elements}. The semimajor axis of the relative
orbit, $a$, is inferred to be 30.8~mas (1.58~AU in linear units). The
orbital period is determined to better than 0.1\% by virtue of the
19.4-yr baseline provided by the velocity measurements (more than 12
cycles).  Perhaps one of the most significant improvements coming from
the combination of the astrometry and spectroscopy is in the
trigonometric parallax. Compared to the value from the HIPPARCOS
catalog, our parallax that takes full account of the orbital motion is
nearly 10~mas smaller, which corresponds to a 50\% change in the
distance to \H\ (see Table~\ref{tab:pm}), and is considerably more
precise. This has important consequences for the luminosity estimates
discussed below.

According to our solution the orbit of the binary is seen nearly
edge-on: the inclination angle is $85\arcdeg \pm 13\arcdeg$. Although
in principle this would allow for eclipses, these are very unlikely
given the 570-day period. Inspection of the HIPPARCOS photometry for
\H\ shows no evidence of any eclipse events (four conjunctions
occurred during the mission), and the peak-to-peak brightness
variation recorded is $\sim$0.1~mag, which is probably at the level of
the uncertainties for a star as faint as $V = 10.80$.  Nevertheless,
carefully scheduled observations may perhaps be of interest since the
satellite measurements do not have the optimal sampling. 

\section{The nature of the companion}
\label{sec:secondary}

The information provided by our combined solution allows a direct
determination of the mass of the unseen companion of \H, given an
estimate of the primary mass. For this we use our effective
temperature determination in \S\ref{sec:spectroscopy} ($T_{\rm eff} =
4600 \pm 150$~K), and the brightness of the object. The latter was
measured during the HIPPARCOS mission as $H_p = 10.9366 \pm 0.0055$,
in the passband of the satellite. Conversion to the Johnson system
\citep{ESA:97} yields $V = 10.80 \pm 0.02$, where the uncertainty is
our conservative estimate that includes the transformation. With our
parallax of $\pi = 19.5 \pm 1.8$~mas (corresponding to a distance of
$51.3 \pm 4.9$~pc) the absolute magnitude in the visual band is $M_V =
7.25 \pm 0.20$, ignoring extinction, in which essentially all of the
uncertainty comes from the parallax error. 

\begin{figure} 
\vskip -1in
\epsscale{0.87} 
\plotone{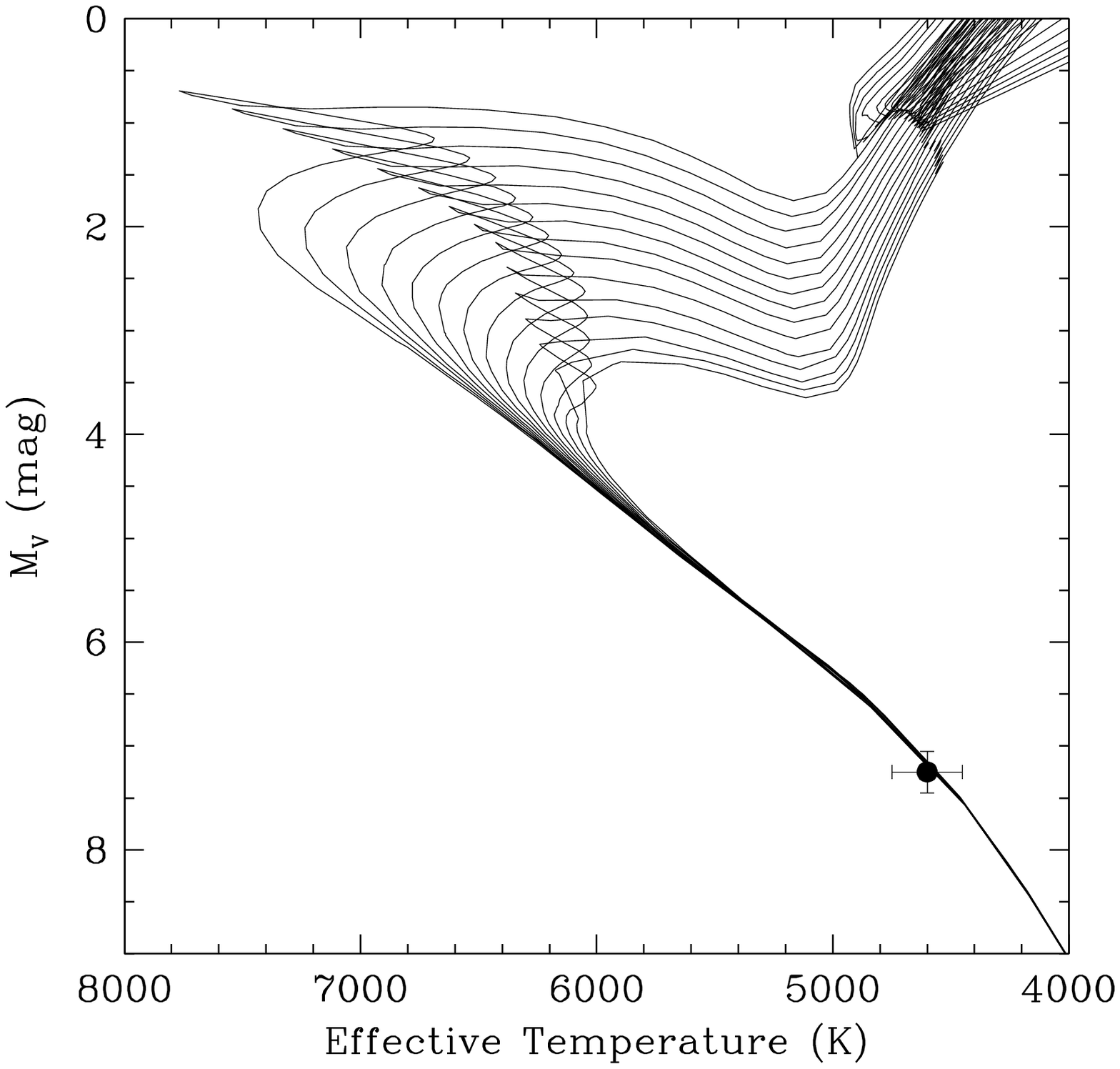}
\vskip -0.5in
 \figcaption[]{Location of \H\ in the H-R diagram, against the
backdrop of model isochrones by \cite{Girardi:00} for solar
metallicity and ages ranging from 1~Gyr to 5~Gyr.\label{fig:hr}}
 \end{figure}

The location of the primary star in the H-R diagram is shown in
Figure~\ref{fig:hr}. Model isochrones from the series by
\cite{Girardi:00} for solar metallicity agree well with the measured
properties of \H, quite independently of age (ranging from 1~Gyr to
5~Gyr in the figure). From these models we estimate the primary mass
to be $M_1 = 0.73 \pm 0.05$~M$_{\sun}$, in which the uncertainty
accounts for observational errors as well as age.  The surface gravity
inferred from the models is $\log g = 4.65 \pm 0.05$, in agreement
with our crude estimate from \S\ref{sec:spectroscopy}.  Although the
precise metallicity of the star is unknown, our tests in
\S\ref{sec:spectroscopy} suggested a composition near solar.
Additionally, the space motion of \H\ in the Galactic frame based on
our parallax, proper motion, and center-of-mass velocity is quite
small ($U = -8$~\kms, $V = -14$~\kms, $W = +9$~\kms, relative to
the Local Standard of Rest), which supports a Population~I origin and
a heavy element abundance probably close to that of the Sun, as we
have assumed. 

The adopted mass for the primary leads to a companion mass of $M_{\rm
comp} = 0.88 \pm 0.05$~M$_{\sun}$. The uncertainty in the primary mass
contributes only about 25\% to this error. The secondary star is thus
20\% more massive than the primary, which implies it cannot be a
single main-sequence star. A more careful examination of our spectra
was carried out using the two-dimensional cross-correlation algorithm
TODCOR \citep{Zucker:94}, to place limits on the brightness of the
companion. This technique often allows the detection and measurement
of faint secondaries that are difficult to see in standard
one-dimensional cross-correlation diagrams. No evidence of another set
of lines was found down to the level of the noise, at about 10\% of
the brightness of the primary. Additionally, \cite{Hemenway:97}
reported an observation of \H\ by speckle interferometry in which no
companions were apparently seen in the range from 30~mas to
0\,\farcs7. Details of this measurement are unavailable.

One possibility is that the companion is a white dwarf. Depending on
its temperature, it may be possible to detect it in the ultraviolet,
although no such observations are available to our knowledge. Given
the present mass of the object, initial-final mass relations for white
dwarfs \citep[e.g.,][and references therein]{Ferrario:05} suggest a
progenitor of approximately 4--5~M$_{\sun}$, corresponding to a mid-B
star. Alternatively, the companion of \H\ could itself be a (closer)
binary composed of lower main-sequence stars, making the system a
hierarchical triple. In this case, however, the mass ratio of the
binary cannot be too small or we would have seen the brighter of the
two stars in our spectra\footnote{A possible exception would be if the
orbital period of the binary is short enough that the stars are spun
up by tidal forces and are rotating very rapidly, in synchronism with
the orbital motion.  In that case the spectral lines may be broad
enough to reduce the contrast and escape detection.}. 

Possible signs of this type of configuration might be seen in the form
of an infrared excess.  $JHK_s$ measurements for \H\ are available
from the 2MASS catalog. With the visual magnitude listed earlier, and
after conversion to the standard Johnson system as defined by
\cite{Bessell:88} using transformations by \cite{Carpenter:01}, we
obtain the observed colors $V\!-J$, $V\!-H$, and $V\!-K$ listed in the
first line of Table~\ref{tab:photometry}.  The predicted colors for a
single star of the assumed mass ($M_1 = 0.73$~M$_{\sun}$) according to
the models by \cite{Girardi:00} for a representative age of 2~Gyr are
indeed bluer, by as much as 0.4~mag.  However, there are a number of
reasons to doubt the theoretical colors in this case, including
indications of missing opacity sources and other limitations in the
physics \citep[e.g.,][and references therein]{Baraffe:98, Delfosse:00,
Chabrier:05}, and especially disagreements with empirical
mass-luminosity relations, particularly in the visual band.
Therefore, we have chosen here to rely on the empirical relations.
Those by \cite{Henry:93} give redder colors than \cite{Girardi:00} for
the primary (see third line of Table~\ref{tab:photometry}), but are
still considerably bluer than observed, by 0.17~mag, 0.29~mag, and
0.25~mag for $V\!-J$, $V\!-H$, and $V\!-K$, respectively. These
differences are 6--7 times larger than the observational errors,
suggesting the infrared excess may be real. 

\begin{figure} 
\epsscale{0.9} 
\plotone{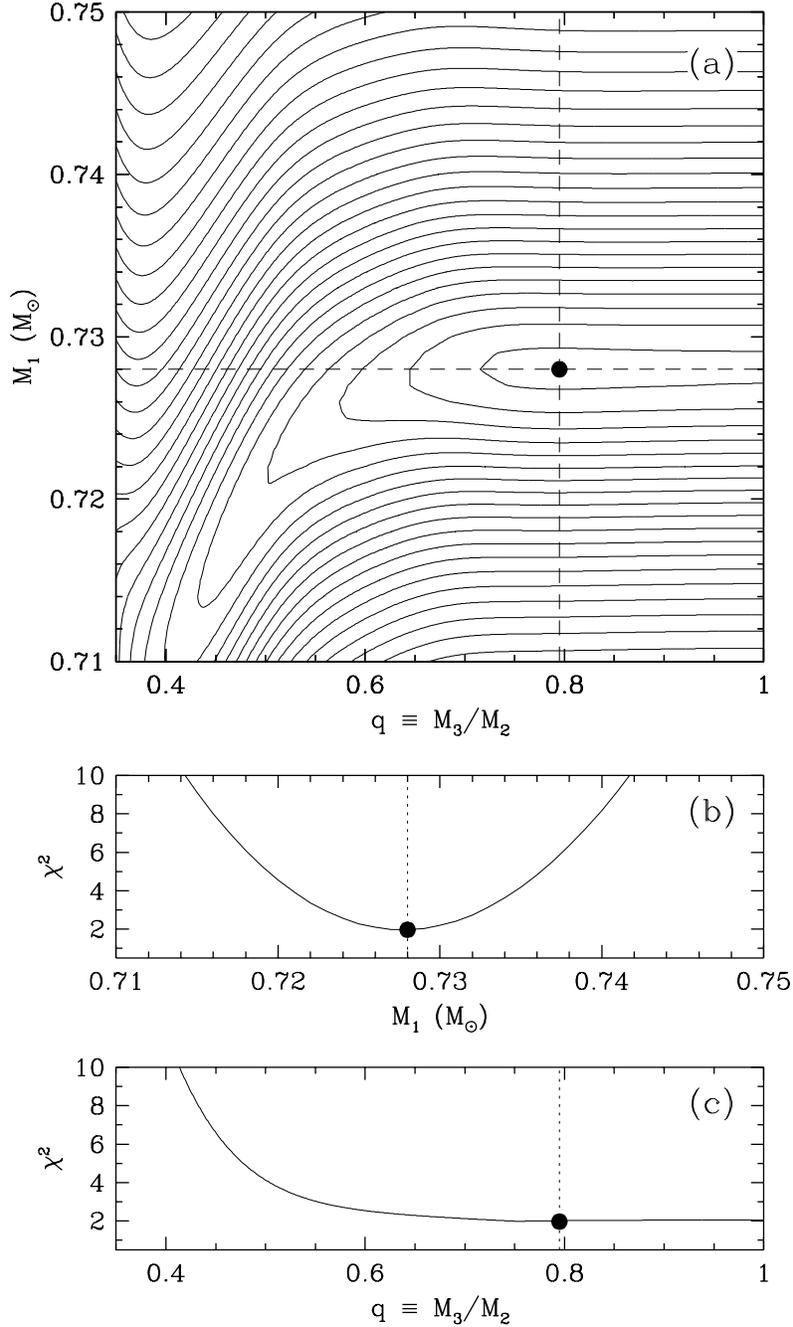}
\vskip 0.8in
 \figcaption[]{Morphology of the $\chi^2$ surface resulting from our
modeling of the observed $V$ and $JHK$ photometry of \H\ with the sum
of three stars. (a) Contours of equal $\chi^2$. The dot represents the
best fit with $M_1 = 0.728$~M$_{\sun}$ and a binary mass ratio of $q =
0.795$, corresponding to $M_2 = 0.49$~M$_{\sun}$ and $M_3 =
0.39$~M$_{\sun}$ (see text); (b) Cross-section along the $M_1$ axis,
showing a well defined minimum; (c) Cross-section along the $q$ axis,
showing a very flat $\chi^2$ surface in the vicinity of the minimum.
Solutions at higher mass ratios and the same $M_1$ are not
significantly worse. \label{fig:contours}}
 \end{figure}

With the aid of the \cite{Henry:93} relations, we have modeled this
excess by computing the visual and $JHK$ magnitudes of the components
of the unseen companion (hereafter star~2 and star~3) over a range of
possible masses for each star, subject to the constraint that their
sum be the value we determined above, $M_{\rm comp} =
0.88$~M$_{\sun}$. We then added the light of these stars to that of
the primary (star~1) in each passband, and calculated the combined
colors for the triple system. For convenience we parameterized the
problem in terms of the mass ratio of the binary ($q = M_3/M_2$, where
star~2 is the more massive component).  Additionally, we allowed for
changes in the mass of the main star, and sought to produce the best
match to the observed colors as well as the combined absolute visual
magnitude, in a $\chi^2$ sense.  A solution was found for a mass $M_1
= 0.728$~M$_{\sun}$ and a binary mass ratio $q = 0.795$, which
reproduces the observed visual brightness and all infrared colors
within the uncertainties (see last row of Table~\ref{tab:photometry}).
Contours of the $\chi^2$ surface as a function of $M_1$ and $q$ are
shown in Figure~\ref{fig:contours}, where the best fit is indicated by
a dot. The masses of the binary components from this model are $M_2 =
0.49$~M$_{\sun}$ and $M_3 = 0.39$~M$_{\sun}$. The predicted magnitude
difference in the visual band between the brighter of these stars and
star~1 is 3.0~mag, which is below our threshold for spectroscopic
detection ($\sim$2.5~mag; see above) and is thus consistent with all
observational constraints. Although this is formally the best fit to
the observations, we note that the $\chi^2$ surface is nearly flat in
the $q$ direction for $q$ larger than about 0.7--0.8 at a fixed value
of $M_1$, so that larger mass ratios up to and including unity are
equally acceptable and reproduce the observed colors just as well (see
Figure~\ref{fig:contours}). The mass of star~1, on the other hand, is
very tightly constrained.  At $q = 1$ stars 2 and 3 would be at their
faintest, each about 3.3~mag fainter than star~1. 

\section{Constrained orbital solution} 
\label{sec:orbit2}

Even though, according to the model above, both stars in the binary
are too faint to be seen individually in our spectra, their combined
light is not negligible compared to the brightness of star~1. The
difference is approximately 2.5~mag in the visual band.  This suggests
that our earlier assumption that the astrometric motion detected by
HIPPARCOS traces only the motion of star~1 may not be entirely
correct. The extra light (even with $\Delta V = 2.5$) must affect the
astrometry at some level, despite our failure to detect the effect
directly, as described in \S\ref{sec:orbit1}. This will cause the
semimajor axis of the photocenter, $a_{\rm phot}$, to be slightly
smaller than that of the primary alone by a factor $(B-\beta)/B
\approx 0.17$, where $B = M_{\rm comp}/(M_1+M_{\rm comp})$ is the mass
fraction of the companion and $\beta = (1+10^{0.4 \Delta V})^{-1}$ is
its fractional light. We therefore adjusted our orbital solution to
allow for this effect, and recomputed all orbital elements. The
results of this ``constrained" solution are seen in the last column of
Table~\ref{tab:elements}.  The semimajor axis of the photocenter is
reduced from $a_{\rm phot} = a_1 = 16.9 \pm 1.5$~mas to $a_{\rm phot}
= 14.9 \pm 1.3$~mas, a small effect. 

\begin{figure} 
\vskip -1.5in
\epsscale{0.9} 
\plotone{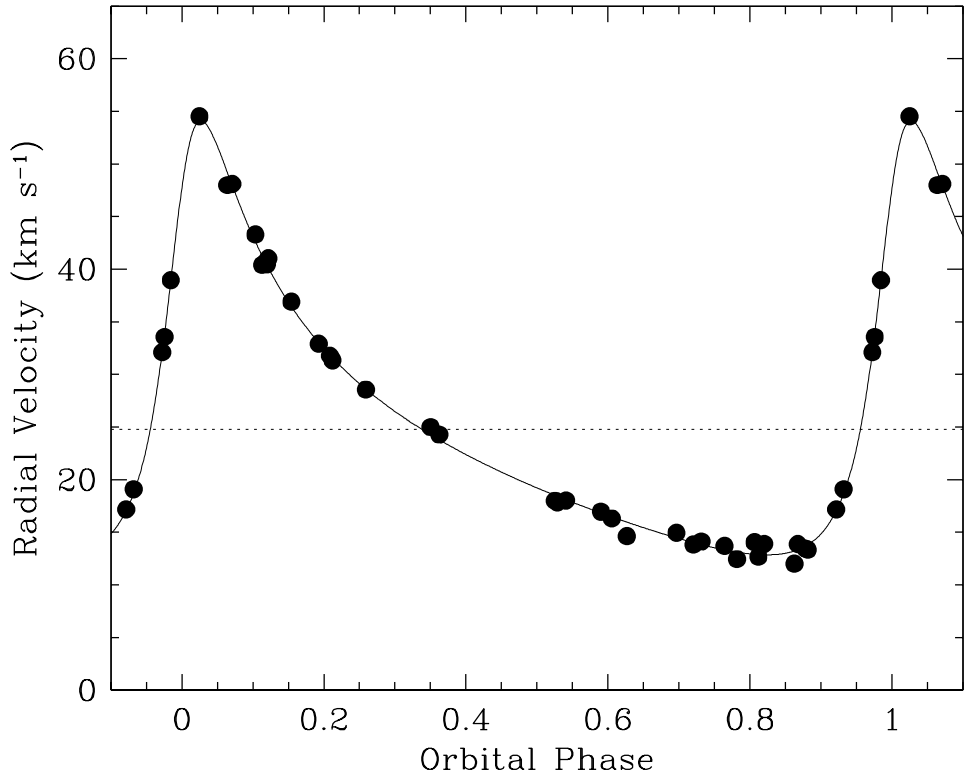}
\vskip -0.5in
 \figcaption[]{Radial velocity observations for \H\ and fitted orbit.
The dotted line represents the center-of-mass velocity of the binary.
Errors are smaller than the size of the points.\label{fig:rvs}}
 \end{figure}

\begin{figure} 
\vskip -1in
\epsscale{0.87} 
\plotone{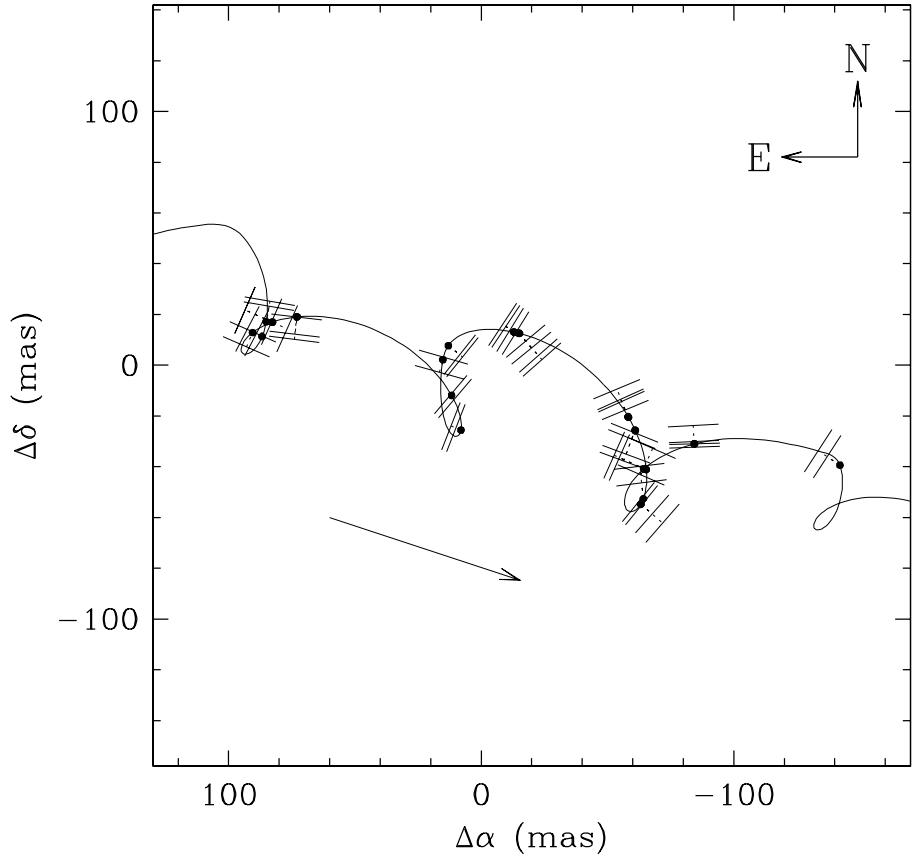}
 \figcaption[]{Path of the center of light of \H\ on the plane of the
sky, along with the HIPPARCOS observations (abscissae residuals). See
text or Figure~\ref{fig:hip2} for an explanation of the graphical
representation of these one-dimensional measurements.  The figure
shows the total motion resulting from the combined effects of
parallax, proper motion, and orbital motion according to the global
solution described in the text.  The arrow indicates the direction and
magnitude of the annual proper motion. \label{fig:hip1}}
 \end{figure}

\begin{figure} 
\vskip -1in
\epsscale{0.87} 
\plotone{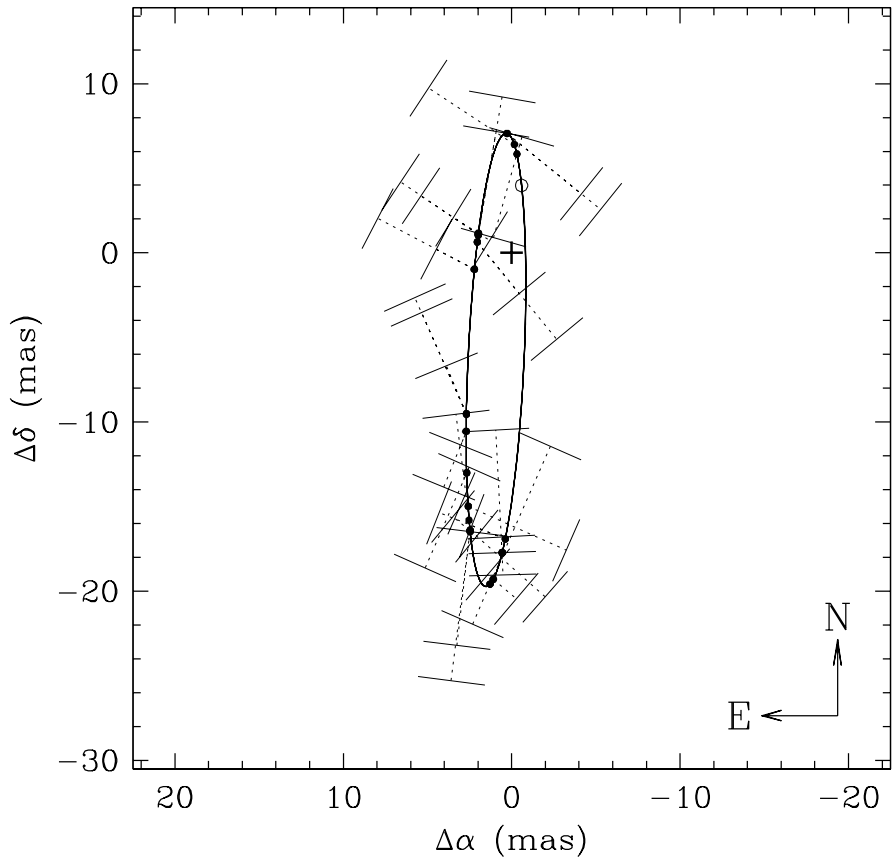}
 \figcaption[]{Residual orbital motion of the center of light of \H\
on the plane of the sky, after removal of the parallactic and
proper-motion components. HIPPARCOS observations are represented as in
Figure~\ref{fig:hip1} (a few measurements with large residuals have
been omitted here for clarity).  Because they are one-dimensional in
nature \citep{ESA:97}, the exact location of each measurement on the
plane of the sky cannot be shown graphically. Filled circles represent
the predicted location on the computed orbit.  The dotted lines
connecting to each filled circle indicate the scanning direction of
the HIPPARCOS satellite for each measurement, and show which side of
the orbit the residual is on. The short line segments at the end of
and perpendicular to the dotted lines indicate the direction along
which the actual observation lies, although the precise location is
undetermined. Occasionally more than one measurement was taken along
the same scanning direction, in which case two or more short line
segments appear on the same dotted lines.  The plus sign in the figure
indicates the center of mass of the binary, and periastron is shown
with an open circle. The direction of motion in the orbit is direct
(counterclockwise).\label{fig:hip2}}
 \end{figure}

The residuals of the radial-velocity observations from this fit are
listed in Table~\ref{tab:rvs}, and those of the HIPPARCOS measurements
are given in Table~\ref{tab:hip}.  Figure~\ref{fig:rvs} displays the
spectroscopic measurements along with the velocity curve as a function
of orbital phase. The astrometric observations on the plane of the sky
are illustrated in Figure~\ref{fig:hip1} and Figure~\ref{fig:hip2}, in
which the axes are parallel to the Right Ascension and Declination
directions. The curious pattern in Figure~\ref{fig:hip1} is the result
of the combined effects of annual parallax, proper motion, and orbital
motion. The dominant contribution is from the proper motion
(80~mas~yr$^{-1}$), which is indicated with an arrow. Parallax and
orbital motion are smaller and comparable effects. In
Figure~\ref{fig:hip2} we have subtracted the proper motion and
parallactic contributions, leaving only the orbital motion with the
570-day period and a semimajor axis of $a_{\rm phot} = 14.9$~mas.  The
direction of motion is direct (counterclockwise).  The individual
HIPPARCOS observations are represented schematically in both of these
figures, but are seen more clearly in Figure~\ref{fig:hip2}.  Because
they are one-dimensional in nature \citep{ESA:97}, the exact location
of each measurement on the plane of the sky cannot be shown
graphically. The filled circles represent the predicted location on
the computed orbit.  The dotted lines connecting to each filled circle
indicate the scanning direction of the HIPPARCOS satellite for each
measurement, and show which side of the orbit the residual is on. The
short line segments at the end of and perpendicular to the dotted
lines indicate the direction along which the actual observation lies,
although the precise location is undetermined. Occasionally more than
one measurement was taken along the same scanning direction, in which
case two or more short line segments appear on the same dotted lines. 

Compared to our previous solution, the parallax is increased slightly
to $\pi = 20.6 \pm 1.9$~mas, which corresponds to a distance of $48.5
\pm 4.5$~pc. The proper motion components change very little
(Table~\ref{tab:pm}).  The binary mass derived with $M_1 =
0.73$~M$_{\sun}$ is marginally larger than before ($M_{\rm comp} =
0.89 \pm 0.06$~M$_{\sun}$). The change in the parallax leads to a
revised absolute magnitude for the system of $M_V = 7.37 \pm 0.20$.
With these updated values of $M_{\rm comp}$ and $M_V$ we repeated the
photometric modeling using the \cite{Henry:93} mass-luminosity
relations, and obtained a solution not very different from the
previous one (the same value of $M_1$, and $q = 0.842$, implying $M_2
= 0.48$~M$_{\sun}$ and $M_3 = 0.41$~M$_{\sun}$).  The brightness
difference between star~1 and the sum of stars 2 and 3 is unchanged,
making another iteration of our orbital solution unnecessary. 

\section{Discussion and concluding remarks}
\label{sec:discussion}

\H\ is an interesting illustration of the complementarity of
spectroscopic and astrometric observations, and in particular of the
utility of the HIPPARCOS intermediate data \citep[see
also][]{Pourbaix:04, Jorissen:04, Pourbaix.et:04}. While the
radial-velocity measurements clearly reveal this object to be a binary
and provide the (single-lined) spectroscopic orbit with a period of
570 days, the combination with the astrometry has allowed us to derive
the dynamical mass of the companion ($M_{\rm comp} = 0.89$~M$_{\sun}$)
and also to correct the seriously biased parallax value from the
original HIPPARCOS reductions. No indications of the companion are
detected in our spectra, yet the orbital solution shows that it is
clearly more massive than the visible star, which is a normal K dwarf
with $M \approx 0.73$~M$_{\sun}$.  While we cannot completely rule out
that the secondary is a massive white dwarf, all the evidence points
quite convincingly toward the conclusion that the companion is itself
a closer binary composed of M dwarfs, making the system a hierarchical
triple. 

The infrared excess that might be expected from such a configuration
appears indeed to be present, as indicated by our modeling of the
measured visual and near-infrared (2MASS) photometry using empirical
mass-luminosity relations by \cite{Henry:93}, along with the
constraint on the total mass of the unseen companion. This modeling
cannot determine the precise mass ratio of the close binary, but is
able to place a lower limit of about $q \approx 0.8$.  Smaller values
would make one of the stars bright enough that it would be seen in our
spectra. We infer masses for these stars of approximately $M_2
=$~0.44---\,0.48~M$_{\sun}$ and $M_3 =$~0.41---\,0.44~M$_{\sun}$. The
period of the close binary is unknown.  We note also that \H\ is
listed as an X-ray source in the ROSAT catalog, with $\log L_X =
28.62$ \citep{Micela:97} and $\log (L_X/L_{\rm bol}) = -2.61$
\citep{Makarov:01}. This X-ray emission might arise naturally from the
presence of the M dwarfs, which are frequently active, particularly if
they are in a short-period configuration so that tidal forces compel
the stars to be in synchronous rotation with the orbital motion. 

Finally, the relative orbit of \H\ and its unseen companion has an
angular semimajor axis of about 33~mas. Given the eccentricity of the
orbit, separations up to about 50~mas are possible at times, which may
permit a direct detection of the secondary with high-resolution
techniques. Alternatively, infrared spectroscopy should be able to
reveal the presence of at least the brighter of the M dwarfs directly,
given the more favorable contrast with the main star at those
wavelengths. 

\acknowledgements 

The author is grateful to P.\ Berlind, M.\ Calkins, D.\ W.\ Latham,
and R.\ P.\ Stefanik for obtaining the spectroscopic observations used
in this work, and to R.\ J.\ Davis for maintaining the CfA echelle
database. B.\ Mason and P.\ Hemenway are thanked for information on
speckle measurements. This paper benefited also from helpful comments
by an anonymous referee. Partial support for this work from NSF grant
AST-0406183 and NASA's MASSIF SIM Key Project (BLF57-04) is
acknowledged.  This research has made use of the SIMBAD database,
operated at CDS, Strasbourg, France, of NASA's Astrophysics Data
System Abstract Service, and of data products from the Two Micron All
Sky Survey, which is a joint project of the University of
Massachusetts and the Infrared Processing and Analysis
Center/California Institute of Technology, funded by NASA and the NSF. 

\appendix 

\section{Incorporating HIPPARCOS data into the global orbital
solution}
\label{appendix}

The intermediate data provided with the HIPPARCOS catalog are the
``abscissae residuals", $\Delta v$, which are the difference between
the satellite measurements (abscissae) along great circles and the
abscissae computed from the 5 standard astrometric parameters. The
standard parameters are the position of the object ($\alpha_0^*$,
$\delta_0$) at the reference epoch $t_0 = 1991.25$, the proper motion
components ($\mu_{\alpha}^*$, $\mu_{\delta}$), and the parallax
($\pi$). We follow here the notation in the HIPPARCOS catalog and
define $\alpha_0^* \equiv \alpha_0 \cos\delta$ and $\mu_{\alpha}^*
\equiv \mu_{\alpha} \cos\delta$, to incorporate the projection
factors.  The goal of an orbital solution making use of the HIPPARCOS
data is to reduce the abscissae residuals below the values obtained
from the 5-parameter solution by taking the orbital motion into
account\footnote{The 5 standard parameters for \H\ were actually
computed as part of a 7-parameter solution incorporating the time
derivatives of the proper motion components (since HIPPARCOS detected
curvature in the proper motion).  Nevertheless, the abscissae
residuals available are always derived from the 5 standard parameters
as listed in the catalog.}.  Following \cite{Pourbaix:00} the $\chi^2$
sum for the abscissae residuals is $\chi^2 = \mbox{\boldmath $\Xi^t~
V^{-1}~ \Xi$}$, where
 \begin{equation}
 \label{eq:xi}
 \mbox{\boldmath $\Xi$} = \mbox{\boldmath $\Delta v$} -
\sum_{k=1}^M {\partial \mbox{\boldmath $v$}\over\partial p_k} \Delta
p_k
 \end{equation}
 and \mbox{\boldmath $\Xi^t$} is the transpose of \mbox{\boldmath
$\Xi$}. 
 In this expression \mbox{\boldmath $\Delta v$} is the array of N
abscissae residuals provided by HIPPARCOS, and $\partial
\mbox{\boldmath $v$}/\partial p_k$ is the array of partial derivatives
of the abscissae with respect to the $k$-th fitted parameter.  The
number $M$ of parameters fitted to the astrometry in the general case
is 12: the 5 standard HIPPARCOS parameters ($p_1 = \alpha_0^*$, $p_2 =
\delta_0$, $p_3 = \mu_{\alpha}^*$, $p_4 = \mu_{\delta}$, $p_5 = \pi$)
and 7 orbital elements ($a_1$, $P$, $e$, $i$, $\omega_1$, $\Omega$,
$T$, represented as $p_k$ with $k=6,...,12$).  \mbox{\boldmath
$V^{-1}$} is the inverse of the covariance matrix of the observations,
containing the abscissae uncertainties and correlation coefficients
\citep[][Vol.\ 3, eqs.\ 17-10 and 17-11]{ESA:97} also provided with
the HIPPARCOS catalog.  Correlations arise because the same original
data were reduced independently by two data reduction consortia
\citep[NDAC and FAST; see][]{ESA:97}, and both results are included in
the solution. 

The partial derivatives $\partial \mbox{\boldmath $v$}/\partial p_k$
for $k=1$ to 5 are given in the HIPPARCOS catalog along with the
abscissae residuals. The remaining derivatives can be expressed in
terms of the partial derivatives of \mbox{\boldmath $v$} with respect
to $\alpha_0^*$ and $\delta_0$. These are \citep[][Vol.\ 3, eq.\
17-15]{ESA:97}
 \begin{equation}
 \label{eq:derivative}
 {\partial\mbox{\boldmath $v$}\over\partial p_k} = {\partial
\mbox{\boldmath $v$}\over\partial\alpha_0^*}{\partial\xi\over\partial
p_k}+{\partial\mbox{\boldmath
$v$}\over\partial\delta_0}{\partial\eta\over\partial
p_k}~~,~~k=6,...,12,
 \end{equation}
 in which $\xi$ and $\eta$ are in our case the rectangular coordinates
of the photocenter relative to the center of mass of the binary on the
plane tangent to the sky at ($\alpha_0^*$, $\delta_0$), which are
given by
 \begin{eqnarray}
\xi & = & \alpha_0^* + \mu_{\alpha}^*(t-t_0) + \pi P_{\alpha} + \Delta X \\
\eta & = & \delta_0  + \mu_{\delta}(t-t_0) + \pi P_{\delta} + \Delta Y~~.
 \end{eqnarray}
 $P_{\alpha}$ and $P_{\delta}$ are the parallactic factors, and the
terms $\Delta X = Bx + Gy$ and $\Delta Y = Ax + Fy$ represent the
orbital motion components, where $A$, $B$, $F$, and $G$ are the
classical Thiele-Innes constants. These depend only on the orbital
elements $a_1$, $i$, $\omega_1$, and $\Omega$ \citep[see,
e.g.,][]{vandeKamp:67}.  $x$ and $y$ are the rectangular coordinates
in the unit orbit given by $x = \cos E - e$ and $y = \sqrt{1-e^2} \sin
E$, with $E$ being the eccentric anomaly. 

As described by \cite{Pourbaix:00}, the nature of the orbital solution
is such that only the derivative corresponding to the semimajor axis
in eq.(\ref{eq:derivative}) needs to be considered. The expression for
\mbox{\boldmath $\Xi$} in eq.(\ref{eq:xi}) then reduces to
 \begin{equation}
 \mbox{\boldmath $\Xi$} = \mbox{\boldmath $\Delta v$} - \sum_{k=1}^5
{\partial \mbox{\boldmath $v$}\over\partial p_k} \Delta p_k -
\left({\partial \mbox{\boldmath $v$}\over\partial\alpha_0^*} \Delta X
+ {\partial \mbox{\boldmath $v$}\over\partial\delta_0} \Delta
Y\right)~. 
 \end{equation}
 That the semimajor axis $a_1$ has actually been eliminated as a
formal adjustable parameter in our case is irrelevant, since it still
appears in the Thiele-Innes constant but is computed from other
elements using eq.(\ref{eq:a1}). 

Finally, the $\chi^2$ for the global solution that combines astrometry
and spectroscopy is computed by adding the term corresponding to the
radial velocities, in the usual manner.

\clearpage

\clearpage

\clearpage

\clearpage

\clearpage

\clearpage

\begin{deluxetable}{cccccc}
\tabletypesize{\scriptsize}
\tablecolumns{6}
\tablewidth{0pt}
\tablecaption{Radial velocity measurements for \H\ in the heliocentric frame.\label{tab:rvs}}
\tablehead{\colhead{HJD}          & \colhead{}     & \colhead{RV}        & \colhead{$\sigma_{\rm RV}$\tablenotemark{a}}  & \colhead{(O$-$C)}  & \colhead{Orbital} \\
           \colhead{(2,400,000+)} & \colhead{Year} & \colhead{($\kms$)}  & \colhead{($\kms$)}           & \colhead{($\kms$)} & \colhead{Phase}}
\startdata 
  46422.9750 &  1985.9767 &   $+$16.31 &   0.31 &  $-$0.06  &  0.605 \\
  46537.6905 &  1986.2907 &   $+$14.05 &   0.57 &  $+$1.11  &  0.806 \\
  46540.7560 &  1986.2991 &   $+$12.68 &   0.41 &  $-$0.25  &  0.812 \\
  46569.7043 &  1986.3784 &   $+$12.01 &   0.73 &  $-$1.43  &  0.862 \\
  52391.6615 &  2002.3180 &   $+$47.99 &   0.86 &  $-$1.38  &  0.064 \\
  52395.7197 &  2002.3291 &   $+$48.10 &   0.59 &  $+$0.03  &  0.071 \\
  52419.6668 &  2002.3947 &   $+$40.40 &   0.70 &  $-$0.96  &  0.113 \\
  52421.6565 &  2002.4002 &   $+$40.59 &   0.50 &  $-$0.30  &  0.116 \\
  52423.6634 &  2002.4056 &   $+$40.47 &   0.60 &  $+$0.05  &  0.120 \\
  52424.6403 &  2002.4083 &   $+$41.03 &   0.70 &  $+$0.83  &  0.122 \\
  52655.0264 &  2003.0391 &   $+$18.00 &   0.71 &  $-$0.54  &  0.525 \\
  52656.9803 &  2003.0444 &   $+$17.85 &   0.50 &  $-$0.60  &  0.529 \\
  52663.9288 &  2003.0635 &   $+$18.03 &   0.56 &  $-$0.09  &  0.541 \\
  52691.9163 &  2003.1401 &   $+$16.96 &   0.54 &  $+$0.09  &  0.590 \\
  52712.8028 &  2003.1973 &   $+$14.63 &   0.97 &  $-$1.37  &  0.627 \\
  52752.7176 &  2003.3066 &   $+$14.95 &   0.60 &  $+$0.44  &  0.697 \\
  52772.7344 &  2003.3614 &   $+$14.13 &   0.71 &  $+$0.26  &  0.732 \\
  52985.0078 &  2003.9425 &   $+$43.29 &   0.51 &  $+$0.53  &  0.104 \\
  53013.9323 &  2004.0217 &   $+$36.91 &   0.44 &  $+$0.36  &  0.154 \\
  53035.8873 &  2004.0818 &   $+$32.92 &   0.51 &  $-$0.19  &  0.193 \\
  53044.9091 &  2004.1065 &   $+$31.77 &   0.47 &  $-$0.14  &  0.209 \\
  53046.9420 &  2004.1121 &   $+$31.34 &   0.47 &  $-$0.32  &  0.212 \\
  53073.7601 &  2004.1855 &   $+$28.56 &   0.44 &  $-$0.12  &  0.259 \\
  53125.7468 &  2004.3278 &   $+$25.00 &   0.44 &  $+$0.67  &  0.350 \\
  53132.7248 &  2004.3470 &   $+$24.29 &   0.44 &  $+$0.45  &  0.362 \\
  53337.0326 &  2004.9063 &   $+$13.83 &   0.46 &  $-$0.25  &  0.720 \\
  53362.0644 &  2004.9749 &   $+$13.70 &   0.46 &  $+$0.32  &  0.764 \\
  53371.9950 &  2005.0020 &   $+$12.45 &   0.39 &  $-$0.72  &  0.782 \\
  53393.8773 &  2005.0620 &   $+$13.90 &   0.39 &  $+$0.97  &  0.820 \\
  53420.9096 &  2005.1360 &   $+$13.88 &   1.13 &  $+$0.47  &  0.867 \\
  53426.8509 &  2005.1522 &   $+$13.46 &   0.51 &  $-$0.27  &  0.878 \\
  53428.8623 &  2005.1577 &   $+$13.35 &   0.51 &  $-$0.52  &  0.881 \\
  53451.7621 &  2005.2204 &   $+$17.17 &   0.36 &  $+$0.11  &  0.921 \\
  53457.7802 &  2005.2369 &   $+$19.09 &   0.50 &  $+$0.37  &  0.932 \\
  53480.8179 &  2005.3000 &   $+$32.11 &   0.41 &  $+$0.03  &  0.972 \\
  53482.6947 &  2005.3051 &   $+$33.57 &   0.44 &  $-$0.26  &  0.976 \\
  53487.7003 &  2005.3188 &   $+$38.96 &   0.44 &  $+$0.07  &  0.984 \\
  53510.6947 &  2005.3818 &   $+$54.52 &   0.43 &  $+$0.13  &  0.025 \\
\enddata
\tablenotetext{a}{Velocity uncertainties include the scale factor described in the text.}
\end{deluxetable}

\clearpage

\begin{deluxetable}{lccc}
\tabletypesize{\scriptsize}
\tablecolumns{4}
\tablewidth{0pc}
\tablecaption{Orbital solutions for \H.\label{tab:elements}}
\tablehead{
\colhead{\hfil~~~~~~~~~~~~~~~~~Parameter~~~~~~~~~~~~~~~~~~} & \colhead{Spectroscopic only} & \colhead{Combined\tablenotemark{a}} & \colhead{Constrained\tablenotemark{a,b}}}
\startdata
\sidehead{Adjusted quantities} \\
\noalign{\vskip -6pt}
~~~~$P$ (days)\dotfill                               &  570.70~$\pm$~0.73\phn\phn     &  570.95~$\pm$~0.52\phn\phn    &  570.98~$\pm$~0.52\phn\phn    \\
~~~~$\gamma$ (\kms)\dotfill                          &  $+24.77$~$\pm$~0.14\phn\phs   &  $+24.86$~$\pm$~0.11\phn\phs  &  $+24.87$~$\pm$~0.11\phn\phs  \\
~~~~$K$ (\kms)\dotfill                               &  20.62~$\pm$~0.28\phn          &  20.75~$\pm$~0.20\phn         &  20.76~$\pm$~0.20\phn         \\
~~~~$e$\dotfill                                      &  0.6103~$\pm$~0.0071           &  0.6110~$\pm$~0.0052          &  0.6110~$\pm$~0.0051          \\
~~~~$\omega_1$ (deg)\dotfill                         &  313.75~$\pm$~0.95\phn\phn     &  314.01~$\pm$~0.76\phn\phn    &  314.02~$\pm$~0.76\phn\phn    \\
~~~~$T$ (HJD$-$2,400,000)\dotfill                    &  52355.2~$\pm$~1.4\phm{2222}   &  52355.0~$\pm$~1.1\phm{2222}  &  52355.0~$\pm$~1.1\phm{2222}  \\
~~~~$i$ (deg)\dotfill                                &  \nodata                       &  85~$\pm$~13                  &  83~$\pm$~15                  \\
~~~~$\Omega$ (deg)\dotfill                           &  \nodata                       &  175.0~$\pm$~8.8\phn\phn      &  179~$\pm$~10\phn             \\
~~~~$\Delta\alpha^*$ (mas)\dotfill                   &  \nodata                       &  $+3.7$~$\pm$~4.0\phs         &  $+3.1$~$\pm$~4.2\phs         \\
~~~~$\Delta\delta$ (mas)\dotfill                     &  \nodata                       &  $-8.7$~$\pm$~1.8\phs         &  $-7.7$~$\pm$~1.7\phs         \\
~~~~$\Delta\mu_{\alpha}^*$ (mas yr$^{-1}$)\dotfill   &  \nodata                       &  $+4.6$~$\pm$~1.4\phs         &  $+4.5$~$\pm$~1.4\phs         \\
~~~~$\Delta\mu_{\delta}$ (mas yr$^{-1}$)\dotfill     &  \nodata                       &  $-3.4$~$\pm$~1.3\phs         &  $-3.2$~$\pm$~1.3\phs         \\
~~~~$\Delta\pi$ (mas)\dotfill                        &  \nodata                       &  $-9.9$~$\pm$~1.8\phs         &  $-8.8$~$\pm$~1.9\phs         \\
\sidehead{Derived quantities} \\						                                                                          
\noalign{\vskip -6pt}								                                                                                          
~~~~$f(M)$ (M$_{\sun}$)\dotfill                      &  0.258~$\pm$~0.011             &  \nodata                      &  \nodata                      \\
~~~~$M_2 \sin i/(M_1+M_2)^{2/3}$ (M$_{\sun}$)\dotfill&  0.6364~$\pm$~0.0092           &  \nodata                      &  \nodata                      \\ 
~~~~$a_1 \sin i$ ($10^6$ km)\dotfill                 &  128.2~$\pm$~1.9\phn\phn       &  \nodata                      &  \nodata                      \\
~~~~$a_1$ (mas)\dotfill                              &  \nodata                       &  16.9~$\pm$~1.5\phn           &  17.9~$\pm$~1.6\phn           \\
~~~~$a_{\rm phot}$ (mas)\dotfill                     &  \nodata                       &  16.9~$\pm$~1.5\phn           &  14.9~$\pm$~1.3\phn\tablenotemark{c}           \\
~~~~$a$ (AU)\tablenotemark{c}\dotfill                &  \nodata                       &  1.580~$\pm$~0.029            &  1.582~$\pm$~0.032            \\
~~~~$a$ (mas)\tablenotemark{c}\dotfill               &  \nodata                       &  30.8~$\pm$~3.2\phn           &  32.5~$\pm$~3.3\phn           \\
~~~~$\mu_{\alpha}^*$ (mas yr$^{-1}$)\dotfill         &  \nodata                       &  $-75.4$~$\pm$~1.4\phn\phs    &  $-75.5$~$\pm$~1.4\phn\phs    \\
~~~~$\mu_{\delta}$ (mas yr$^{-1}$)\dotfill           &  \nodata                       &  $-25.0$~$\pm$~1.3\phn\phs    &  $-24.8$~$\pm$~1.3\phn\phs    \\
~~~~$\pi$ (mas)\dotfill                              &  \nodata                       &  19.5~$\pm$~1.8\phn           &  20.6~$\pm$~1.9\phn           \\
\enddata
\tablenotetext{a}{Incorporates radial velocities, HIPPARCOS measurements (abscissae residuals), and the Tycho-2 proper motions.}
\tablenotetext{b}{Assumes the light from the secondary is not negligible (a brightness difference of $\Delta V = 2.5$ mag); see text.}
\tablenotetext{c}{Assumes a value for the primary mass of $M_1 = 0.73$~M$_{\sun}$.}
\end{deluxetable}

\clearpage

\begin{deluxetable}{cccccccc}
\tabletypesize{\scriptsize}
\tablewidth{0pt}
\tablecaption{HIPPARCOS abscissae residuals for \H.\label{tab:hip}}
\tablehead{\colhead{HJD} & \colhead{} & \colhead{$v$} &
\colhead{$\sigma_v$} & \colhead{} & \colhead{Corr.} & \colhead{$O-C$} &
\colhead{Orbital} \\
\colhead{(2,400,000+)} & \colhead{Year} & \colhead{(mas)} &
\colhead{(mas)} & \colhead{Cons.\tablenotemark{a}} & \colhead{Coef.\tablenotemark{b}} &
\colhead{(mas)} & \colhead{Phase}}
\startdata 
    47884.3184 &  1989.9776 &   $-$6.60 &   4.60  & F &   0.762 &  $-$5.24  &    0.170 \\
    47884.1723 &  1989.9772 &   $-$8.65 &   5.40  & N &   0.762 &  $-$7.30  &    0.170 \\
    47901.6313 &  1990.0250 &   +1.61 &   5.67  & F &   0.531 &  $-$6.77  &    0.200 \\
    47901.5947 &  1990.0249 &   +5.55 &   4.55  & N &   0.531 &  $-$2.84  &    0.200 \\
    48020.3010 &  1990.3499 &   +3.81 &   4.31  & F &   0.736 &  +0.01  &    0.408 \\
    48020.1914 &  1990.3496 &   $-$2.70 &   5.51  & N &   0.736 &  $-$6.49  &    0.408 \\
    48046.1241 &  1990.4206 &  +20.01\phn &   8.67  & F &   0.868 & +12.78\phn  &    0.453 \\
    48046.0146 &  1990.4203 &  +20.01\phn &  10.75\phn  & N &   0.868 & +12.78\phn  &    0.453 \\
    48046.3068 &  1990.4211 &   +1.86 &   4.72  & F &   0.773 &  $-$5.37  &    0.454 \\
    48046.3798 &  1990.4213 &   +8.67 &   5.82  & N &   0.773 &  +1.44  &    0.454 \\
    48069.0619 &  1990.4834 &   +6.63 &   8.84  & F &   0.878 &  +0.25  &    0.494 \\
    48069.0619 &  1990.4834 &   $-$0.14 &   9.83  & N &   0.878 &  $-$6.53  &    0.494 \\
    48069.2445 &  1990.4839 &   $-$2.29 &   6.59  & F &   0.000 &  $-$8.62  &    0.494 \\
    48203.1086 &  1990.8504 &   $-$1.80 &   3.97  & F &   0.789 &  +2.00  &    0.728 \\
    48203.0721 &  1990.8503 &   $-$4.01 &   4.66  & N &   0.789 &  $-$0.22  &    0.728 \\
    48249.2397 &  1990.9767 &   $-$3.38 &  11.29\phn  & F &   0.950 &  $-$3.45  &    0.809 \\
    48249.2397 &  1990.9767 &   $-$1.29 &  12.87\phn  & N &   0.950 &  $-$1.37  &    0.809 \\
    48367.3615 &  1991.3001 &  $-$17.97\phn &   5.69  & F &   0.798 &  $-$1.14  &    0.016 \\
    48367.2520 &  1991.2998 &  $-$24.12\phn &   7.29  & N &   0.798 &  $-$7.32  &    0.016 \\
    48383.7978 &  1991.3451 &   $-$7.19 &   6.53  & F &   0.853 &  $-$5.25  &    0.045 \\
    48383.7978 &  1991.3451 &   $-$8.67 &   7.31  & N &   0.853 &  $-$6.71  &    0.045 \\
    48454.3275 &  1991.5382 &   $-$1.37 &   6.17  & F &   0.853 &  $-$0.54  &    0.168 \\
    48454.2910 &  1991.5381 &   +1.18 &   7.44  & N &   0.853 &  +2.00  &    0.168 \\
    48454.8024 &  1991.5395 &   +4.31 &   3.93  & F &   0.768 &  +5.72  &    0.169 \\
    48454.7658 &  1991.5394 &   +2.88 &   4.82  & N &   0.768 &  +4.26  &    0.169 \\
    48458.3088 &  1991.5491 &  $-$17.61\phn &   4.83  & F &   0.825 & $-$12.14\phn  &    0.175 \\
    48458.3453 &  1991.5492 &  $-$19.79\phn &   5.64  & N &   0.825 & $-$14.34\phn  &    0.175 \\
    48458.7471 &  1991.5503 &  $-$10.35\phn &   4.28  & F &   0.753 &  $-$4.42  &    0.176 \\
    48458.7471 &  1991.5503 &  $-$13.81\phn &   4.59  & N &   0.753 &  $-$7.92  &    0.176 \\
    48551.9954 &  1991.8056 &   $-$6.38 &   5.31  & F &   0.799 &  $-$7.26  &    0.339 \\
    48551.9223 &  1991.8054 &   $-$5.40 &   5.89  & N &   0.799 &  $-$6.29  &    0.339 \\
    48552.2145 &  1991.8062 &   $-$2.14 &   6.75  & F &   0.683 &  $-$2.89  &    0.340 \\
    48552.3606 &  1991.8066 &  $-$11.07\phn &  10.06\phn  & N &   0.683 & $-$11.80\phn  &    0.340 \\
    48562.6972 &  1991.8349 &   +5.40 &   6.78  & F &   0.779 &  +9.22  &    0.358 \\
    48562.5146 &  1991.8344 &   +8.30 &   9.05  & N &   0.779 & +12.09\phn  &    0.358 \\
    48562.9164 &  1991.8355 &   $-$1.35 &   5.10  & F &   0.730 &  +2.59  &    0.358 \\
    48562.9894 &  1991.8357 &   $-$4.05 &   6.93  & N &   0.730 &  $-$0.10  &    0.359 \\
    48595.8984 &  1991.9258 &   $-$9.20 &   9.97  & F &   0.894 & $-$10.50\phn  &    0.416 \\
    48595.7889 &  1991.9255 &  $-$11.43\phn &  11.07\phn  & N &   0.894 & $-$12.75\phn  &    0.416 \\
    48629.5380 &  1992.0179 &  $-$17.27\phn &   4.36  & F &   0.778 &  $-$7.63  &    0.475 \\
    48629.7206 &  1992.0184 &  $-$23.85\phn &   4.97  & N &   0.778 & $-$14.20\phn  &    0.475 \\
    48638.9614 &  1992.0437 &   +5.21 &   4.58  & F &   0.735 & +11.59\phn  &    0.492 \\
    48638.8519 &  1992.0434 &   $-$0.26 &   4.85  & N &   0.735 &  +6.12  &    0.491 \\
    48639.2901 &  1992.0446 &   $-$5.36 &   4.10  & F &   0.686 &  +0.77  &    0.492 \\
    48639.3632 &  1992.0448 &   $-$7.20 &   4.02  & N &   0.686 &  $-$1.06  &    0.492 \\
    48760.4436 &  1992.3763 &   +9.38 &   9.65  & F &   0.751 & +10.30\phn  &    0.704 \\
    48760.4070 &  1992.3762 &   $-$2.08 &  13.41\phn  & N &   0.751 &  $-$1.18  &    0.704 \\
    48808.2183 &  1992.5071 &  +10.67\phn &  11.83\phn  & F &   0.875 &  +8.18  &    0.788 \\
    48809.0218 &  1992.5093 &   +4.26 &  13.06\phn  & N &   0.875 &  +1.84  &    0.789 \\
    48809.2410 &  1992.5099 &   +3.03 &   5.48  & F &   0.505 &  +0.89  &    0.790 \\
    48809.0949 &  1992.5095 &   +1.71 &   9.46  & N &   0.505 &  $-$0.42  &    0.790 \\
    48942.7033 &  1992.8753 &   $-$2.01 &   6.43  & F &   0.777 &  $-$9.08  &    0.024 \\
    48942.7033 &  1992.8753 &   +2.19 &   7.92  & N &   0.777 &  $-$4.88  &    0.024 \\
\enddata
\tablenotetext{a}{Data reduction consortium responsible for the measurement: F = FAST, N = NDAC \citep[see][]{ESA:97}.}
\tablenotetext{b}{Correlation coefficient between the FAST and NDAC abscissae taken on the same great circle.}
\end{deluxetable}

\clearpage

\begin{deluxetable}{lccc}
\tablecolumns{4}
\tablewidth{0pc}
\tablecaption{Parallax and proper motion components of \H.\label{tab:pm}}
\tablehead{
\colhead{} & \colhead{$\mu_{\alpha}^*$} & \colhead{$\mu_{\delta}$} & \colhead{$\pi$} \\
\colhead{\hfil~~~~~~~~~~~~~~~~~~~~Source~~~~~~~~~~~~~~~~~~~~~} & \colhead{(mas~yr$^{-1}$)} & \colhead{(mas~yr$^{-1}$)} & \colhead{(mas)}
}
\startdata
HIPPARCOS catalog \citep{ESA:97}\dotfill & $-$80.0~$\pm$~2.4\phs\phn & $-$21.6~$\pm$~1.9\phs\phn & 29.4~$\pm$~2.7 \\
Initial fit (this paper)\tablenotemark{a}\dotfill         & $-$76.1~$\pm$~2.2\phs\phn & $-$24.6~$\pm$~1.8\phs\phn & 19.4~$\pm$~1.9 \\
Tycho-2 catalog \citep{Hog:00}\dotfill   & $-$74.9~$\pm$~1.8\phs\phn & $-$25.3~$\pm$~1.8\phs\phn & \nodata \\
Combined fit (this paper)\tablenotemark{b}\dotfill        & $-$75.4~$\pm$~1.4\phs\phn & $-$25.0~$\pm$~1.3\phs\phn & 19.5~$\pm$~1.8 \\
Constrained fit (this paper)\tablenotemark{c}\dotfill     & $-$75.5~$\pm$~1.4\phs\phn & $-$24.8~$\pm$~1.3\phs\phn & 20.6~$\pm$~1.9 \\
\enddata
\tablenotetext{a}{Incorporates radial velocities and HIPPARCOS abscissae residuals, but not the Tycho-2 proper motions.}
\tablenotetext{b}{Includes the Tycho-2 proper motions, and assumes the companion contributes no light.}
\tablenotetext{c}{Accounts for the small light contribution from the companion.}
\end{deluxetable}

\clearpage

\begin{deluxetable}{lcccc}
\tabletypesize{\scriptsize}
\tablewidth{0pc}
\tablecaption{Photometry for \H.\label{tab:photometry}}
\tablehead{
\colhead{} & \colhead{$M_V$} & \colhead{$V-J$} & \colhead{$V-H$} & \colhead{$V-K$} \\
\colhead{\hfil~~~~~~~~~~~~~~~~~~~~~~~~Source~~~~~~~~~~~~~~~~~~~~~~~~} & \colhead{(mag)} & \colhead{(mag)} & \colhead{(mag)} & \colhead{(mag)}
}
\startdata
Observed ($\pi$ from combined solution)\dotfill     & 7.25~$\pm$~0.20  &  2.32~$\pm$~0.03  &  2.99~$\pm$~0.04  &  3.10~$\pm$~0.04  \\
\cite{Girardi:00} isochrones\tablenotemark{a}\dotfill     & 7.25             &  1.96             &  2.60             &  2.67             \\
\cite{Henry:93}\tablenotemark{a}\dotfill                  & 7.22             &  2.15             &  2.70             &  2.85             \\
Observed ($\pi$ from constrained solution)\dotfill  & 7.37~$\pm$~0.20  &  2.32~$\pm$~0.03  &  2.99~$\pm$~0.04  &  3.10~$\pm$~0.04  \\
Model for triple system\dotfill                            & 7.14             &  2.34             &  2.95             &  3.12             \\
\enddata
\tablenotetext{a}{Predicted photometry for a single star with $M_1 = 0.73$~M$_{\sun}$.}
\end{deluxetable}

\clearpage

\end{document}